\documentclass[twocolumn,pra,superscriptaddress,noshowpacs,epsf]{revtex4}
\usepackage{amssymb}
\usepackage{amsmath}
\usepackage{graphicx}

\begin{document}
\title{Variable-frequency-controlled coupling in charge qubit
circuits: Effects of microwave field on qubit-state readout}
\date{\today}
\author{Xiao-Ling He}
\affiliation{Department of Physics and Surface Physics Laboratory (National Key
Laboratory), Fudan University, Shanghai 200433, China}
\affiliation{Frontier Research System, The Institute of Physical and Chemical Research
(RIKEN), Wako-shi 351-0198, Japan}
\author{Yu-xi Liu}
\affiliation{CREST, Japan Science and Technology Agency (JST),
Kawaguchi, Saitama 332-0012, Japan} \affiliation{Frontier
Research System, The Institute of Physical and Chemical Research
(RIKEN), Wako-shi 351-0198, Japan}
\author{J. Q. You}
\affiliation{Department of Physics and Surface Physics Laboratory (National Key
Laboratory), Fudan University, Shanghai 200433, China}
\affiliation{Frontier Research System, The Institute of Physical and Chemical Research
(RIKEN), Wako-shi 351-0198, Japan}
\author{Franco Nori}
\affiliation{Frontier Research System, The Institute of Physical
and Chemical Research (RIKEN), Wako-shi 351-0198, Japan}
\affiliation{CREST, Japan Science and Technology Agency (JST),
Kawaguchi, Saitama 332-0012, Japan} \affiliation{Center for
Theoretical Physics, Physics Department, Center for the Study of
Complex Systems, The University of Michigan, Ann Arbor, MI
48109-1040, USA}

\begin{abstract}
To implement quantum information processing, microwave fields
are often used to manipulate superconuducting qubits. We study
how the coupling between superconducting charge qubits can be
controlled by variable-frequency magnetic fields. We also study
the effects of the microwave fields on the readout of the
charge-qubit states. The measurement of the charge-qubit states
can be used to demonstrate the statistical properties of
photons.

\end{abstract}


\maketitle

\section{Introduction}

Superconducting quantum circuits are good candidates for
implementing quantum information processing~\cite{you1,makhlin}. To
construct universal quantum computing, controllable couplings
between any pair of qubits are required. Theoretical methods for
switchable couplings in charge-qubit circuits have been proposed
by changing the amplitude of the bias magnetic flux, e.g., in
Refs.~\cite{makhlin,you2,you3}. However, in experiments, it is
much easier to produce fast and precise {\it frequency} shifts
of the radio-frequency (RF) control signals, as opposed to
changing the {\it amplitude} of the dc signal. Methods using
variable-frequency-controlled couplings in superconducting
flux-qubit circuits have been studied~\cite{liu1} theoretically,
comparing with the coupling approach using the dressed
states~\cite{devoret,liu,sahin}. In this scheme, the two qubits
can be coupled to (or decoupled from) each other by modulating
the frequencies~\cite{liu1} of externally applied
variable-frequency magnetic fields to match (or mismatch) the
combination of frequencies of the two qubits. The coherent
oscillations and conditional gate operations of two
superconducting charge qubits with always-on coupling have been
demonstrated~\cite{pashkin} experimentally. Therefore, the next
step for charge qubits would be to design superconducting
quantum circuits with switchable couplings.

Here, we first generalize our approach~\cite{liu1} using the
variable-frequency-controlled coupling in {\it flux} qubit
circuits to the {\it charge} qubit circuit proposed in
Ref.~\cite{you3}. This proposal has the following advantages:
(i) the coupling between different charge qubits can be
implemented fast by changing the frequency of the externally
applied classical field; (ii) these proposed charge qubits
always work at their optimal points, and thus the qubits are
mostly immune from  charge noise~\cite{voin}, produced by
uncontrollable charge fluctuations; (iii) no additional circuit
is needed to realize this controllable coupling.

Besides the controllable coupling, measuring the qubit state is
also a very important step in quantum information processing. In
superconducting quantum circuits, microwave fields are often
used to implement quantum information processing. Here we focus
on how microwave fields affect the readouts of the qubit states.
In particular, we explore the effect of quantized fields with
different statistical properties on measurement results of the
qubit states when the charge qubits are placed inside a
microcavity, e.g. a three-dimensional cavity~\cite{you4,liu2} or
a superconducting transmission line~\cite{blais}.

The paper is organized as follows: in Sec.~II, we generalize the
variable-frequency-controlled coupling approach in flux-qubit
circuits~\cite{liu1} to that in charge-qubit
circuits~\cite{you3}. In Sec.~III, we study the effect of the
classical and quantized microwave fields on the readout of the
qubit states. In Sec.~IV, we compare the classical and quantum
treatment of the large Josephson junction. Finally, conclusions
are presented in Sec.~V.

\section{Hamiltonian with variable-frequency controlled couplings}

We first very briefly review the model Hamiltonian proposed in
Ref.~\cite{you3} for two coupled superconducting charge qubits
by sharing a large Josephson junction (JJ) (see Fig.~1). The
large JJ is classically treated and its charge energy $E_{c0}$
is neglected.  The dc biased magnetic field $\Phi_{e}$ is
externally applied through the area between the large JJ and the
first qubit. Each qubit is also biased by a dc voltage $V_{Xi}$
via the gate capacitance $C_{i}$ ($i=1,\,2$). The Hamiltonian of
the superconducting circuit is~\cite{you3}
\begin{eqnarray}
H\!&\!=\!&\!\sum_{i=1}^{2}\left[ E_{i}(V_{Xi})-2E_{Ji}\cos
\left(\frac{\pi \Phi _{e}}{\Phi _{0}}-\frac{\gamma
}{2}\right)\cos \varphi
_{i}\right]\nonumber\\
&&\!-E_{J0}\cos \gamma  \label{H1}
\end{eqnarray}
with $E_{i}(V_{Xi})=E_{ci}(n_{i}-C_{i}V_{Xi}/2e)^{2}$. Here
$E_{ci}=2e^2/(C_{i}+2C_{Ji})$ and $E_{Ji}$ are the charge and
Josephson energies of the $i$th charge qubit. $E_{J0}$ is the
Josephson energy of the large JJ. The number $n_{i}$ of excess
Cooper pairs in the superconducting island is canonically
conjugate to the average phase drop
$\varphi_{i}=(\varphi_{iA}+\varphi_{iB})/2$ of the $i$th charge
qubit. The phase drop across the large JJ is $\gamma$.
Considering that the critical current $I_{0}\equiv 2\pi
E_{J0}/\Phi _{0}$ of the large JJ is much larger than the
critical currents $I_{ci}\equiv 2\pi E_{Ji}/\Phi_{0}$ of the
charge qubits, the phase $\gamma$ across the large JJ is very
small. We can expand the functions of the phase drop $\gamma$ in
Eq.~(\ref{H1}) into a series and and retain the terms up to
second order in the parameters $\eta _{i}=(I_{ci}/I_{0}) <1$. In
this case, Eq.~(\ref{H1}) can be reduced to
\begin{equation}\label{eq:2}
H=\sum_{i=1}^{2}\left[\varepsilon_{i}(V_{Xi})\sigma_{z}^{(i)}
-\bar{E}_{Ji}\sigma_{x}^{(i)}\right] -\chi _{12}
\sigma _{x}^{\left( 1\right)}\sigma _{x}^{\left( 2\right) }
\end{equation}
in the spin-$\frac{1}{2}$ representation based on the charge 
states $|0\rangle\equiv |\!\uparrow\rangle$ and 
$|1\rangle\equiv |\!\downarrow\rangle$ that correspond to 
zero and one excess Cooper pairs in each Cooper-pair box,
where $\varepsilon_{i}(V_{Xi})=\frac{1}{2}E_{ci}(C_iV_{Xi}/e -1)$,
and
\begin{eqnarray*}
\bar{E}_{Ji} &\mathbf{=}&E_{Ji}\cos \left( \frac{\pi \Phi
_{e}}{\Phi _{0}} \right) \left[ 1-\frac{3}{8}\sin ^{2}(\frac{\pi
\Phi _{e}}{\Phi _{0}})\left( \eta _{i}^{2}+3\eta _{j}^{2}\right)
\right],
\end{eqnarray*}
with $\ i,\, j=1,\, 2\,(i\neq j)$. The coupling constant
$\chi_{12}$ between the two charge qubits is
\begin{equation}
\chi_{12}\mathbf{=}L_{J}I_{c1}I_{c2}\sin ^{2}\left(\frac{\pi \Phi _{e}}{\Phi _{0}%
}\right),  \label{coupling strength}
\end{equation}
where $L_{J}=\Phi _{0}/2\pi I_{0}$ is the Josephson inductance
of the large JJ. It is clear that the coupling between the two
qubits is realized via this effective inductance.

\begin{figure}
\includegraphics[width=3.2in,  
bbllx=100,bblly=562,bburx=475,bbury=741]{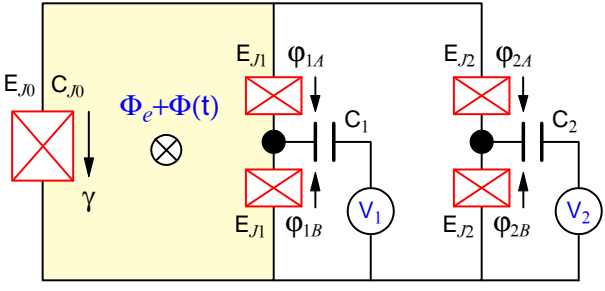} 
\caption[]{(Color online)~Schematic diagram of two charge qubits
coupled by a (left) large Josephson junction (JJ) with coupling
energy $E_{J0}$ and capacitance $C_{J0}$. For the $i$th charge
qubit (where $i=1,2$), a superconducting island (denoted by a
filled circle) is connected to two identical small JJs (each
with coupling energy $E_{Ji}$ and capacitance $C_{Ji}$). Also,
this island is biased by the voltage $V_i=V_{Xi}+V_{g}^{(i)}(t)$ via
a gate capacitance $C_i$, where $V_{Xi}$ is a static (dc) gate
voltage and $V_{g}^{(i)}(t)$ is a time-dependent (ac) microwave gate
voltage. Moreover, a static (dc) magnetic flux $\Phi_e$ plus a
microwave-field-induced magnetic flux $\Phi(t)$ (ac) are applied
to the (yellow) region between the large JJ and the first charge
qubit. }\label{fig1}
\end{figure}

Now, we study how to apply our variable-frequency-controlled
approach~\cite{liu1} to the above charge-qubit
circuits~\cite{you3}. We assume that besides the dc voltages
$V_{Xi}$ and the dc magnetic flux $\Phi_{e}$, an ac microwave
voltage $V_g^{(i)}(t)=V_{gi}\cos \left(\omega _{gi}t\right)$ with the
frequency $\omega _{gi}$ is applied to the superconduction
island of the $i$th qubit via its gate capacitance, and an
additional variable-frequency (ac) magnetic flux  $\Phi\left(
t\right) =\Phi _{c}\sin \left(\omega t\right)$ is also applied
through the area between the large JJ and the first charge qubit
(see Fig.~\ref{fig1}). To make our proposed charge-qubit more
immune from the uncontrollable charge fluctuations, it is also
assumed that two charge qubits work at their optimal points,
i.e. the applied dc voltages $V_{Xi}$ satisfy the condition
$\varepsilon_{i}(V_{Xi})=0$. Considering these conditions, 
the Hamiltonian in Eq.~(\ref{eq:2}) becomes
\begin{eqnarray}
H \!&\!=\!&\! \sum_{i=1}^{2}\left[-\bar{E}_{Ji}\sigma
_{x}^{(i)}+\varepsilon _{0}^{(i)}\cos(\omega _{gi}t) \sigma
_{z}^{(i)}\right] -\chi _{12}\sigma _{x}^{(1)}\sigma _{x}^{(2)}
\nonumber \\
&&\!+\left[ g_{12}\sigma _{x}^{(1)}\sigma
_{x}^{(2)}-\sum_{i=1}^{2}(g_{i}\sigma _{x}^{(i)})\right] \sin(\omega
t) \label{eq:4},
\end{eqnarray}
where $\varepsilon _{0}^{(i)} =E_{ci}(C_{i}V_{gi}/2e)$ and $g_{i}=2
E_{Ji}\sin \left(\pi \Phi _{e}/\Phi _{0}\right) \xi$. The
parameters $g_{12}$ and $\xi$ are given by
\begin{eqnarray*}
g_{12} &=&L_{J}I_{c1}I_{c2}\sin \left( \frac{2\pi \Phi
_{e}}{\Phi _{0}}\right) J_{1}(\varphi _{c}),
\end{eqnarray*}
and
\begin{eqnarray*}
\xi &=&J_{1}(\varphi _{c})\left[ 1-\frac{3(\eta _{i}^{2}+3\eta
_{j}^{2})}{16} \left( 1-\cos \left(\frac{2\pi \Phi _{e}}{\Phi
_{0}}\right)J_{0}(\varphi _{c})\right)
\right] \\
&+&\frac{3}{8}\cos ^{2}\left( \frac{\pi \Phi _{e}}{\Phi
_{0}}\right) J_{0}(\varphi _{c})J_{1}(\varphi _{c})(\eta
_{i}^{2}+3\eta _{j}^{2}).
\end{eqnarray*}
Here $\varphi_{c}=2\pi \Phi _{c}/\Phi _{0}$ and $J_{n}$ is the
$n$th-order Bessel function of the first kind.

In the rotating reference frame at the frequency $\omega _{gi}$
about $\sigma _{x}^{(i)}$, the Hamiltonian in Eq.~(\ref{eq:4})
is rewritten as
\begin{eqnarray}
H \!&\!=\!&\!\sum_{i=1}^{2}\left[ (\hbar \omega _{gi}-\bar{E}_{Ji})\sigma
_{x}^{(i)}+\varepsilon _{0}^{(i)}\sigma _{z}^{(i)}\right] -\chi _{12}\sigma
_{x}^{(1)}\sigma _{x}^{(2)} \nonumber\\
&&\!+\left[ g_{12}\sigma _{x}^{(1)}\sigma
_{x}^{(2)}-\sum_{i=1}^{2}g_{i}\sigma_{x}^{(i)}\right]\sin(\omega
t).\label{eq:5}
\end{eqnarray}
To eliminate the $\sigma _{x}^{(i)}$ term in Eq.~(\ref{eq:5}),
the frequency $\omega _{gi}$ of the microwave field applied to
the gate capacitance is set as $\hbar \omega _{gi}\simeq
\bar{E}_{Ji}$. Furthermore, we can tune the flux $\Phi_{e}$ so
that the coupling strength $\chi_{12}$ is less than the coupling
strength $g_{12}$. Also, we tune the gate voltage $V_{gi}$ so
that the large detuning condition $\left\vert \varepsilon
_{0}^{(2)}-\varepsilon _{0}^{(1)}\right\vert=\Delta\gg \chi _{12}$
can be satisfied. Under this condition, the always-on
interaction $\chi_{12}$ is negligibly small, and the Hamiltonian
in Eq.~(\ref{eq:5}) is reduced~\cite{quality factor} to
\begin{eqnarray}
H \approx \sum_{i=1}^{2}\hbar\omega_{i}\sigma _{z}^{(i)}+\left[
g_{12}\sigma _{x}^{(1)}\sigma
_{x}^{(2)}-\sum_{i=1}^{2}g_{i}\sigma
_{x}^{(i)}\right]\sin(\omega t),\label{eq:H_I_pm}
\end{eqnarray}%
with $\hbar\omega_{1}= \varepsilon _{0}^{(1)}-\chi ^{^{\prime }}$,
$\hbar\omega_{2}=\varepsilon _{0}^{(2)}+\chi ^{\prime }$, and $ \chi
^{\prime}=\chi _{12}^{2}/2\Delta$.

Let us discuss how the interaction between two qubits can be
switched on and off via Eq.~(\ref{eq:H_I_pm}) by changing the
frequency $\omega$ of the variable-frequency magnetic flux
$\Phi(t)=\Phi_{c}\sin(\omega t)$. Equation~(\ref{eq:H_I_pm})
shows that the two qubits are approximately decoupled from each
other when there is no applied ac magnetic flux $\Phi(t)$.
However if the frequency $\omega$ of $\Phi(t)$ is tuned to
satisfy the condition $\omega =\omega_1+\omega_2$, then two qubits can be
simultaneously flipped by the variable-frequency magnetic flux
via the interaction Hamiltonian
\begin{equation}
V_{I}=g_{12}\sigma _{-}^{(1)}\sigma _{-}^{(2)}\
+g_{12}^{*}\sigma _{+}^{(1)}\sigma _{+}^{(2)} \, ,  \label{eq:7}
\end{equation}
where the contributions of other fast oscillating terms are
negligibly small. If the frequency $\omega$ of $\Phi(t)$
satisfies the condition $\omega =\omega_2-\omega_1$, then one qubit
can be flipped by another with the help of the
variable-frequency magnetic flux through the interaction
Hamiltonian
\begin{equation}
V_{I}^{\prime}=g_{12}\sigma _{+}^{(1)}\sigma _{-}^{(2)}\
+g_{12}^{*}\sigma _{-}^{(1)}\sigma _{+}^{(2)},\label{eq:8}
\end{equation}
after neglecting other fast oscillating terms.

A single-qubit operation can also be implemented via the
variable-frequency magnetic flux $\Phi(t)$. For example, if
$\omega =\omega_1$, or $\omega=\omega_2$, then the first or second
qubit can be selectively rotated around the $x$ axis. When there
is no variable-frequency magnetic flux, a rotation around the
$z$-axis can be implemented for each qubit. Therefore, any logic
gate (see, e.g., Ref.~\onlinecite{DBE}) can be realized by using 
single-qubit operations and
two-qubit operation via the Hamiltonians in Eqs.~(\ref{eq:7})
and (\ref{eq:8}).

\section{Effect of microwave fields on supercurrents in the measurement
of qubit states}

Above, we have shown that the interaction between the two qubits
can be switched on and off using a variable-frequency magnetic
flux. Two-qubit operations can be implemented, and entangled
states between two qubits can also be generated, using
Eq.~(\ref{eq:7}) or ~(\ref{eq:8}). To implement the readout of
two-qubit states, we need to calculate the circulating
supercurrent $\hat{I}$ contributed by the two qubits~\cite{you3}. 
The operator of the supercurrent $\hat{I}$ of the two qubits is
given by
\begin{eqnarray}
\hat{I} \!&\!=\!&\!\sin \left(\frac{\pi \Phi _{e}}{\Phi
_{0}}\right)\left( I_{c1}\sigma
_{x}^{(1)}+I_{c2}\sigma _{x}^{(2)}\right)  \\
&&\! - \frac{1}{4I_{0}}\sin \left(\frac{2\pi \Phi _{e}}{\Phi
_{0}}\right)\left[ I_{c1}^{2}+I_{c2}^{2}+2I_{c1}I_{c2}\,\sigma
_{x}^{(1)}\sigma _{x}^{(2)}\right].\notag \label{eq:H_I_delta}
\end{eqnarray}
For any given state (e.g., $|\Psi\rangle$) of two qubits, the
supercurrent can be obtained by
\begin{equation}
I=\langle \Psi| \hat{I} |\Psi\rangle.
\end{equation}
Note that two-qubit operations are always related to the
microwave fields. The supercurrent $I$ might be different for
different microwave fields with different statistical
properties. Below, we study how the different microwave fields
affect the supercurrent $I$.

\subsection{Classical microwave field}

We now focus on two-qubit entangled states, created from the
ground state $|g_{1},g_{2}\rangle $ via the two-qubit
interaction Hamiltonian in Eq.~(\ref{eq:7}). For these created
entangled two-qubit states, the contribution of the average
values of single-qubit operators $\sigma_{x}^{(i)}$ ($i=1,\,2$)
to the supercurrent is zero, and the supercurrent $I$ is only
determined by the two-qubit operator $\sigma_{x}^{(1)}\sigma
_{x}^{(2)}$ as follows
\begin{equation}
\langle \hat{I}\rangle =-\,\frac{\eta I_{c}}{2}\sin \left( \frac{2\pi \Phi _{e}%
}{\Phi _{0}}\right) \left( 1+\left\langle \sigma
_{x}^{(1)}\sigma _{x}^{(2)}\right\rangle \right).\ \label{eq:10}
\end{equation}
Here, for simplicity, the two qubits are supposed to have
identical critical supercurrent $I_{c1}=I_{c2}=I_{c}$, and then
$\eta _{1}=\eta _{2}=\eta$. The quantum fluctuation of the total
supercurrent $\hat{I}$ is
\begin{equation}
\overline{( \triangle \hat{I}) ^{2}}=\langle \hat{I}^{2}\rangle
-\langle \hat{I}\rangle ^{2},
\end{equation}
which can be further given by
\begin{eqnarray}
\overline{( \triangle \hat{I})^{2}} \!&\!=\!&\!\frac{\eta ^{2}}{4}
I_{c}^{2}\sin ^{2}\left( \frac{2\pi \Phi _{e}}{\Phi _{0}}\right)
\left[1-\left(\left\langle \sigma _{x}^{(1)}
\sigma _{x}^{(2)}\right\rangle\right)^{2}\right]\nonumber \\
&&\!+2I_{c}^{2}\sin ^{2}\left( \frac{\pi \Phi _{e}}{\Phi
_{0}}\right) \left( 1+\left\langle \sigma _{x}^{(1)}\sigma
_{x}^{(2)}\right\rangle \right).\label{eq:11}
\end{eqnarray}
Considering that the ratio $\eta$ is small, the first term in
Eq.~(\ref{eq:11}) can be neglected in the following
calculations. In this case, the supercurrent fluctuation has a
similar behavior to the supercurrent $\langle \hat{I}\rangle $
of Eq.~(\ref{eq:10}) and $\overline{(\triangle \hat{I})
^{2}}\propto \langle \hat{I}\rangle$, when entangled two-qubit
states are created from the ground state $|g_{1},g_{2}\rangle$
through the Hamiltonian in Eq.~(\ref{eq:7}). For convenience, we
define a reduced quantity $\kappa $ to describe the supercurrent
$\langle \hat{I}\rangle $ and supercurrent fluctuation
$\overline{( \triangle \hat{I}) ^{2}}$ as
\begin{eqnarray}
\kappa (\tau)&=&1+\left\langle \sigma _{x}^{(1)}\sigma
_{x}^{(2)}\right\rangle=\frac{\overline{( \triangle \hat{I})
^{2} }}{2I_{c}^{2}\sin ^{2}(\pi \Phi _{e}/\Phi
_{0})}\nonumber\\
&=&\frac{-2\langle \hat{I }\rangle }{\eta I_{c}\sin (2\pi \Phi
_{e}/\Phi _{0})}\, . \label{general fluctuation}
\end{eqnarray}

If two charge qubits are initially in an entangled state $\cos
\theta |g_{1},g_{2}\rangle +\sin \theta e^{i\phi
}|e_{1},e_{2}\rangle $ and the evolution of the  two qubits is
governed by the Hamiltonian in Eq.~(\ref{eq:7}), then the
reduced supercurrent or supercurrent fluctuation $\kappa (\tau)$
can be given by
\begin{equation}
\kappa_{c}=1+\sin (2\theta )\cos \left(\tau \right) \cos \phi
+\cos (2\theta )\sin \left(\tau \right),
\end{equation}
which means that $\kappa _{c}$ is an ac signal. Here,
$\tau=|g_{12}|t$,  with the evolution time $t$. If the initial
state is $\cos \theta |g_{1},g_{2}\rangle +\sin \theta
|e_{1},e_{2}\rangle $, then $\kappa _{c}=1+\sin (\tau +2\theta
)$. When the evolution time $\tau _{0}=-2\theta +2n\pi -\frac{\pi
}{2}$, $\kappa _{c}=0$, which gives rise to  
$\left\langle \sigma _{x}^{(1)}\sigma_{x}^{(2)}\right\rangle=-1$. 
Thus, in this case both the total supercurrent and the 
supercurrent fluctuation become zero.

\subsection{Quantized microwave field}

Now let us consider the case when the variable-frequency
magnetic flux $\Phi _{c}\cos (\omega t)$ is replaced by a
quantized magnetic flux, $\Phi _{q}a^{+}+\Phi _{q}^{\ast }a$,
with frequency $\omega =\omega_1+\omega_2$.
Following the same way as the above derivation of
Eq.~(\ref{eq:7}), we can obtain an interaction Hamiltonian
$H_{I}$ between the quantized magnetic flux and the two charge
qubits

\begin{equation}
H_{I}=\xi _{12}\;a^{+}\;\sigma _{-}^{(1)}\sigma _{-}^{(2)}\ +\xi
_{12}^{\ast}\;a\;\sigma _{+}^{(1)}\sigma _{+}^{(2)}\ \
\label{eq:15}
\end{equation}%
where
\begin{equation}
\xi _{12}=-\,\frac{2\pi\Phi _{q}L_{J}I_{c1}I_{c2}}{\Phi _{0}}
\sin \left(\frac{2\pi\Phi_{e}}{\Phi_{0}}\right).
\end{equation}
This model indicates that one photon can flip both qubits
simultaneously.

We now consider that the two qubits are initially in the state
$\cos \theta \,|g,g\rangle +\sin \theta\, e^{i\phi }|e,e\rangle
$ and the quantum field is initially in a state $\sum
D(n)\,|n\rangle$, here $D(n)$ will be given below for a given
state. From the Hamiltonian~(\ref{eq:15}), the total system
evolves to
\begin{eqnarray}
\Psi (\tau )&=&\sum_{n=0}^{n=\infty }\left[ a_{n}(\tau
)\,|e,e,n\rangle+b_{n}(\tau )\,|g,g,n+1\rangle \right]\nonumber\\
&+&\cos \theta \,D(0)\,|g,g,0\rangle, \label{wave}
\end{eqnarray}
where
\begin{eqnarray*}
a_{n}(\tau ) &=&
\cos (\tau \sqrt{n+1}\ )\sin \theta \,e^{i\phi}\,D(n) \\
&-&\sin (\tau \sqrt{n+1}\ )\cos \theta \,D(n+1), \\
b_{n}(\tau ) &=&
\cos (\tau \sqrt{n+1}\ )\cos \theta \,D(n+1) \\
&+&\sin (\tau \sqrt{n+1}\ )\sin \theta\, e^{i\phi }\,D(n),
\end{eqnarray*}%
with the rescaled dimensionless time $\tau =\left\vert \xi
_{12}\right\vert t$. Using Eq.~(\ref{general fluctuation}) and
Eq.~(\ref{wave}),  at the time $\tau$, the reduced supercurrent
expectation value or supercurrent fluctuation $\kappa_{q}(\tau)$
in the case of the quantized field is
\begin{equation}
\kappa_{q}(\tau)=1+2\,{\rm Re}\left\{w_{0}(\tau )\cos \theta\,
D(0)+\sum_{n=0}^{\infty }\left[ u_{n}(\tau )\,v_{n}(\tau
)\right] \right\}\, ,  \label{Iq}
\end{equation}
where
\begin{eqnarray*}
w_{0}(\tau ) &=&\cos \tau \sin \theta \,e^{-i\phi }\,D^{\ast
}(0)-\sin \tau \cos\theta\, D^{\ast }(1), \\
u_{n}(\tau ) &=&
\cos (\tau \sqrt{n+2})\sin \theta \,e^{-i\phi }\,D^{\ast }(n+1) \\
&-&\sin (\tau \sqrt{n+2})\cos \theta \,D^{\ast }(n+2), \\
v_{n}(\tau ) &=&
\cos (\tau \sqrt{n+1})\cos \theta\, D(n+1) \\
&+&\sin (\tau \sqrt{n+1})\sin \theta\, e^{i\phi}\,D(n).
\end{eqnarray*}%
Equation~(\ref{Iq}) shows that the supercurrent expectation
value $\langle \hat{I}\rangle$ consists of a dc component
$-(\eta I_{c}/2)\sin( 2\pi \Phi _{e}/\Phi _{0})$ and different
ac components, which are modulated by time-dependent factors,
e.g. $\cos (\tau \sqrt{n+1})$.

\begin{figure}
\includegraphics[width=3.2in,  
bbllx=25,bblly=158,bburx=510,bbury=669]{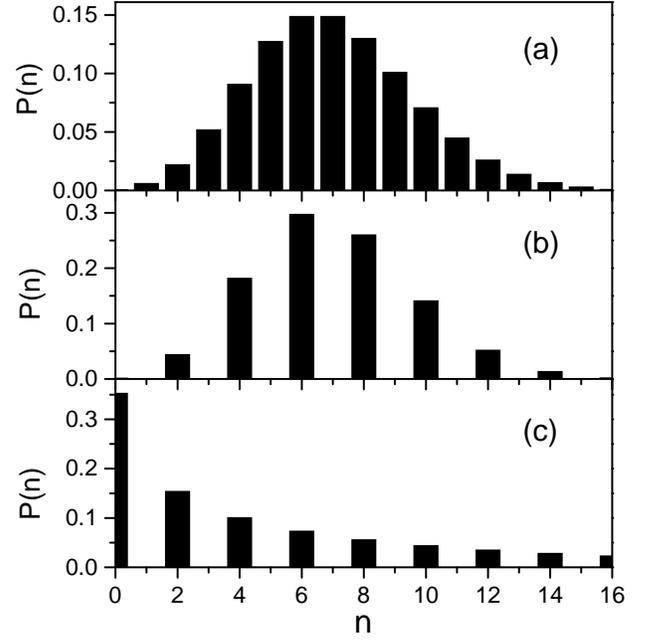} 
\caption{The photon number distribution $P(n)$ with average
photon number $\bar{n}=7$\ of (a) coherent state
$|\protect\alpha \rangle $; (b) superposition of coherent states
$(|\protect\alpha \rangle +|-\protect\alpha \rangle )/N_{+}$;
and (c) squeezed vacuum state $|0,\protect\zeta \rangle
$.}\label{fig2}
\end{figure}

We further specify that the quantized field is initially in
several different quantum states~\cite{book Quantum Optics},
e.g., (i) the coherent state
\begin{equation}
|\alpha \rangle =e^{-\bar{n}/2}\sum \frac{\alpha
^{n}}{\sqrt{n!}}\,|n\rangle\, ,
\end{equation}
 with $\alpha
=\sqrt{\bar{n}}e^{i\varphi }$;  (ii) the superposition of two
coherent states
\begin{equation}
|\alpha_{s}\rangle=\frac{1}{N_{+}}(|\alpha \rangle +|-\alpha
\rangle )=\sum \frac{\alpha ^{2n}}{\sqrt{(2n)!\cosh
\bar{n}}}|2n\rangle\, ,
\end{equation}
with $N_{+}=\sqrt{2(1+e^{-2\bar{n}})}$; and (iii) the squeezed
vacuum state
\begin{equation}
|0,\zeta \rangle =\sum \frac{\sqrt{(2n)!}}{n!\sqrt{\cosh
r}}[-e^{i\beta }\tanh (r/2)]^{n}|2n\rangle\, ,
\end{equation}
with the squeezing parameter $\zeta \equiv re^{i\beta }$ and
$\bar{n}=\sinh ^{2}r$. Here, $\bar{n}$ is the average photon
number. The photon number distributions $P(n)$ (e.g.,
$P(n)=|D(n)|^2=|\langle n|\alpha\rangle|^2$ for a coherent state) 
of the above three states are shown in Fig.~\ref{fig2}. Physically,
coherent states display Poissonian distribution and the
fluctuations of both quadrature components are equal to the
standard quantum fluctuation limit, $\Delta X_{1}=\Delta
X_{2}=1/2$. Squeezed states have sub-poisson distribution and
the fluctuation for one of the quadrature components can be
squeezed, e.g., $\Delta X_{1}<1/2$.

If the qubits are initially in the ground state
$|g_{1},g_{2}\rangle $ and the quantized field is initially in
the coherent state $|\alpha \rangle$, the reduced supercurrent
expectation value or the supercurrent fluctuation is obtained
from Eq.~(\ref{Iq}):
\begin{equation}\label{eq:18}
\kappa _{q}(\tau )=1-2\cos \varphi\,
e^{-\bar{n}}\sum_{n=0}^{\infty } A(n)\sin (\tau
\sqrt{n+1})\,\cos (\tau \sqrt{n}),
\end{equation}
with $A(n)=\bar{n}^{(n+\frac{1}{2})}/(n!\sqrt{n+1})$.
Equations~(\ref{Iq})  and ~(\ref{eq:18}) show that the
supercurrent expectation value $\langle\hat{I}\rangle$ and the
supercurrent fluctuation $\overline{(\Delta\hat{I}\rangle)^2}$
are very sensitive to the phase $\varphi $ of the coherent
state. If $\varphi =\pi/2$, $\kappa_{q}$ has only a dc
component; however, when $\varphi\neq \pi/2$, $\kappa_{q}$
consists of many different ac components.

If the qubits are initially in the ground state
$|g_{1},g_{2}\rangle$, but the quantized fields are initially in
the squeezed vacuum states or superposition of coherent states,
then from Eq.~(\ref{Iq}),  $\kappa_{q}(\tau)$ is given by
\begin{equation}
\kappa _{q}(\tau )=1-2\sum_{n=0}^{\infty }{\rm Re}\left[
B(n)\sin (\tau \sqrt{n+1})\,\cos (\tau \sqrt{n})\right]\, ,
\end{equation}
with $B(n)=D^{\ast}(n+1)\,D(n)$. Because of $D(2n+1)=0$, 
there are $B(n)=0$ and $\kappa _{q}(\tau)=1$. Thus, 
the oscillatory evolution disappears.

\begin{figure}
\includegraphics[width=3.2in,  
bbllx=36,bblly=125,bburx=538,bbury=669]{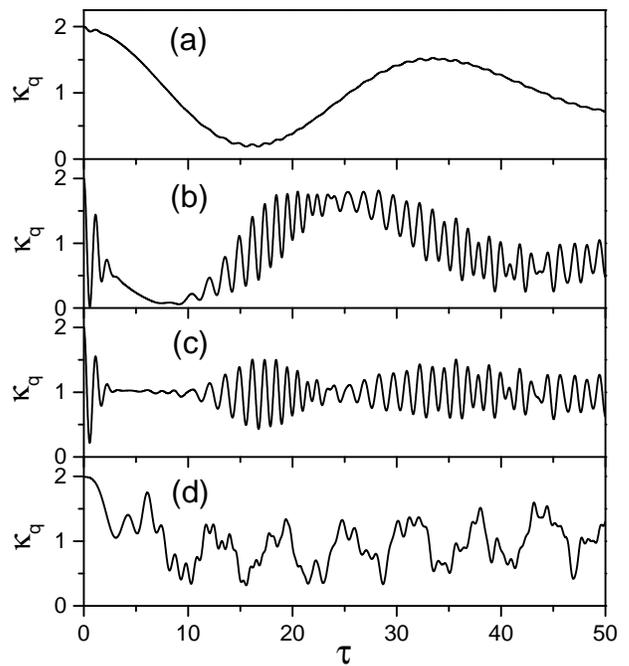} 
\caption{Evolution $\protect\kappa _{q}(\protect\tau )$ of the
reduced total supercurrent expectation and reduced supercurrent
fluctuation from the initial qubit state $(|g,g\rangle
+|e,e\rangle )/\protect\sqrt{2}$ in the presence of the quantum
field initially in (a) coherent state $|\protect\alpha \rangle $
with the phase $\protect\varphi =\protect\pi /2$; (b) coherent
state $\protect\varphi =0$; (c) superposition of coherent states
$(|\protect\alpha \rangle +|-\protect\alpha \rangle )/N_{+}$
with $\protect\varphi =0$; (d) squeezed vacuum state
$|0,\protect\zeta \rangle $. The irregularity of oscillations
originates from the interference effect of the photon component
of the above states.}\label{fig3}
\end{figure}

From Eq.~(\ref{general fluctuation}), we know that the
macroscopic supercurrent expectation value $\langle
\hat{I}\rangle $ can be described by $\kappa_{q}$.
Figure~(\ref{fig3}) shows that the supercurrent of the charge
qubits are different  with the same initial qubit state
$(|g,g\rangle+|e,e\rangle)/\sqrt{2}$  but with different initial states
of the quantum field. From Eq.~(\ref{Iq}), in the case of the
coherent state $|\alpha \rangle $ with the phase $\varphi =\pi
/2$, the total supercurrent $\langle \hat{I}\rangle $ displays a
sinusoidal-like evolution, as shown in Fig.~\ref{fig3}(a).
However, when $\varphi =0$, the total supercurrent is shown in
Fig.~\ref{fig3}(b). If the quantized field is initially in a
superposition of coherent states,  the total supercurrent
$\langle \hat{I}\rangle $, as shown in Fig.~\ref{fig3}(c),
demonstrates the collapse and partial-revival phenomena. In the
case of the squeezed vacuum state,  the total supercurrent
approximately displays an ac current with a quasi-periodic
evolution, which is demonstrated by Fig.~\ref{fig3}(d).  All
irregular oscillations of the supercurrent expectation or
supercurrent fluctuation reflect the coherent interference that
comes from the coherent superpositions of the different photon
number states. 
The different initial photon states result in
different output of the measurement of the charge-qubit states.
Therefore, the measurement of the charge-qubit states can
demonstrate the statistical properties of the photons, and
charge qubits could be served as photon detectors.

\section{Quantization treatment on large Josephson junction}

In the Hamiltonian~(\ref{H1}), the Josephson energy term
$E_{c0}N{}^{2}$ of the large JJ is neglected and the large JJ
acts as an effective inductance $L_{J}$~\cite{you3,Grajcar}. We
now consider a quantum mechanical treatment for the large JJ.
Considering the additional term of charge energy
$E_{c0}N{}^{2}$, the Hamiltonian of the large JJ can be written
as
\begin{equation}
H_{0}=E_{c0}\,N^{2}-E_{J0}\cos \gamma,
\end{equation}
with the charge energy $E_{c0}$ and the excess Cooper pairs $N$.
Because the large JJ works in the phase regime, the spectrum of
the large JJ is approximately equivalent to a harmonic
oscillator $H_{0}=\hbar \omega _{p}a^{\dagger }a$, with the
plasma frequency
\begin{equation}
\omega _{p}=\frac{1}{\hbar }\sqrt{8E_{J}^{(0)}E_{c}^{(0)}}.
\end{equation}
The bosonic operators $a$ and $a^{\dagger}$ are defined by
\begin{equation}
a=\frac{\varsigma}{2}\gamma +i\frac{1}{2\varsigma}N,\,\,\,\,
a^{\dagger}=\frac{\varsigma}{2}\gamma -i\frac{1}{2\varsigma}N\,,
\end{equation}
and the phase drop $\gamma $ is expressed as
\begin{equation}
\gamma =\frac{1}{\varsigma}(a^{\dagger }+a)\, ,
\end{equation}
with $\varsigma =(E_{J}^{(0)}/2E_{c}^{(0)})^{1/4}$. Due to the
large critical supercurrent of the large JJ, one can expand the
phase drop $\gamma $ in Eq.~(\ref{H1}) into a series and retain terms to the
first order of $\gamma$. Finally, a spin-boson interaction
between the two charge qubits and the large JJ is achieved:
\begin{eqnarray}
H \!&\!=\!&\!\sum_{i=1}^{2}\left[ \varepsilon _{i}(V_{xi})\sigma _{z}^{\left(
i\right) }-E_{Ji}\cos \left( \frac{\pi \Phi _{e}}{\Phi _{0}}\right) \sigma
_{x}^{\left( i\right) }\right] \nonumber \\
&&\!+\hbar \omega _{p}a^{\dagger }a+\sum_{i=1}^{2}\left[
g_{i0}\sigma _{x}^{(i)}(a^{+}+a)\right]\, ,
\end{eqnarray}%
where $g_{i0}=-\left(E_{Ji}/2\varsigma \right) \sin \left( \pi
\Phi _{e}/\Phi _{0}\right)$. We assume that the plasma frequency
$\omega_{p}$ of the large JJ is much larger than the splitting
of the qubits. Thus the large JJ is always in the ground state
when the qubits are operated. Following the standard technique
of adiabatic elimination~\cite{quality factor}, we can eliminate
the bosonic mode of the large JJ and obtain an effective
interaction Hamiltonian between the two qubits: $\chi
_{12}\sigma _{x}^{\left( 1\right) }\sigma _{x}^{\left( 2\right)
}$,  with the coupling strength $\chi _{12}=-2g_{10}g_{20}/\hbar
\omega _{p}$. Using the expression of $g_{10}$, $g_{20}$, and
$\omega_{p}$, one can easily confirm that this inter-qubit
coupling is the same as that in Eq.~(\ref{eq:2}). Here the large
JJ serves as the data bus to virtually mediate the interaction
between the two qubits. Therefore, the classical and quantum
treatment to the large JJ are equivalent to each other.
Generalizing the two-qubit system to the multi-qubit system, the
effective inter-qubit coupling term reads $\sum_{i>j}\chi
_{ij}\sigma _{x}^{\left( i\right) }\sigma _{x}^{\left( j\right)
}$ with $\chi _{ij}=-2g_{i0}g_{j0}/\hbar \omega _{p}$.  Because
$g_{i0}=-\left(E_{Ji}/2\varsigma \right) \sin \left( \pi \Phi
_{e}/\Phi _{0}\right)$, the coupling $\chi _{ij}$ is tunable by
changing the static magnetic field $\Phi_{e}$ applied to the
loop. We should point out that if the dc magnetic flux $\Phi
_{e}$ is replaced by an ac variable-frequency magnetic flux
$\Phi_{e}(t)$, then the qubit can be selectively coupled to the
data-bus by a well-chosen frequency-matching condition between
the qubit, data bus, and the variable-frequency magnetic flux.

\section{Conclusions}

In summary, we have studied a variable-frequency-control
approach in charge-qubit circuits: the switchable coupling
between the two charge qubits can be implemented by changing the
frequency of the externally applied magnetic flux. Single-qubit
operations can also be addressed and operated selectively. The
charge qubits are chosen to work at their optimal points, so the
effect of the noise, resulted from uncontrollable charge
fluctuations, on the charge qubits is much suppressed. Moreover,
the effects of the microwave field on the supercurrent of the
two qubits are discussed. It is found that the supercurrent of
the qubits significantly depends on the states of the microwave
field. We also discuss the quantum treatment of the large JJ and
find that both quantum and classical treatments are equivalent
to each other. If the two-qubit circuit is generalized to many
qubits, the interaction $\sigma^{(i)}_{x}\sigma^{(j)}_{x}$ can
also be achieved.

\begin{acknowledgments}
F. N. was supported in part by the US National Security Agency
(NSA), Army Research Office (ARO), Laboratory of Physical
Sciences (LPS), and the National Science Foundation grant No.
EIA-0130383. X. L. H. and J. Q. Y. were supported by the SRFDP
and the National Natural Science Foundation of China 
grant Nos.~10534060 and~ 10625416.
\end{acknowledgments}


\end{document}